%% file: aaskaii_final.tex
\documentclass[a4paper,11pt]{article}
\usepackage{aaskaiid}

\usepackage{xcolor}
\usepackage{textgreek}
\usepackage[utf8]{inputenc}
\usepackage[english]{babel}

\usepackage{hyperref}
\hypersetup{
    unicode, 
    colorlinks=true,
    linkcolor=linkcolor,
    citecolor=linkcolor,
    filecolor=linkcolor,
    urlcolor=linkcolor,
}
\usepackage{color,colortbl}
\definecolor{linkcolor}{rgb}{0.0,0.3,0.5}
\usepackage{tensind}
\tensordelimiter{?}
\DeclareGraphicsExtensions{.bmp,.png,.jpg,.pdf}
\usepackage{verbatim}
\usepackage[normalem]{ulem}
\usepackage{orcidlink}
\usepackage{soul}


\title{Understanding Pulsar Wind Nebulae with the SKA}
\ShortTitle{PWNe with the SKA}

\author[1,2,3]{Joseph D. Gelfand \orcidlink{0000-0003-4679-1058}}
\ShortName{Gelfand et al.} 
\author[4]{C.-Y. Ng\orcidlink{0000-0002-5847-2612}}
\author[5,6]{B. Posselt\orcidlink{0000-0003-2317-9747}}
\author[]{Mallory S. E. Roberts \orcidlink{0000-0002-9396-9720}}
\author[7]{Subir Bhattacharyya\orcidlink{0000-0002-7891-2136}}
\author[8]{Shi Dai\orcidlink{0000-0002-9618-2499}}
\author[9]{Rene Breton \orcidlink{0000-0001-8522-4983}}
\author[9]{Benjamin Stappers \orcidlink{0000-0001-9242-7041}}
\author[10]{Andrea Possenti\orcidlink{0000-0001-5902-3731}}
\author[11]{Jason Hessels \orcidlink{0000-0003-2317-1446}}
\author[4]{Yifan Sun \orcidlink{0000-0001-5874-1432}}
\author[1,2]{Moaz Abdelmaguid\orcidlink{0000-0002-4441-7081}}

\affiliation[1]{NYU Abu Dhabi, Science Division (Physics Program), Abu Dhabi, UAE}
\emailAdd{jg168@nyu.edu}
\affiliation[2]{Center for Astrophysics and Space Science, NYU Abu Dhabi, Abu Dhabi, UAE}
\affiliation[3]{Center for Cosmology and Particle Physics (affiliate member), New York University, New York, NY, USA}

\emailAdd{ncy@astro.physics.hku.hk}
\affiliation[4]{Department of physics, The University of Hong Kong, Pokfulam Rd, HKSAR}

\emailAdd{bettina.posselt@physics.ox.ac.uk}
\affiliation[5]{Department of Astrophysics, University of Oxford, Denys Wilkinson Building, Keble Road, Oxford, OX1 3RH, UK}
\affiliation[6]{Department of Astronomy \& Astrophysics, Pennsylvania State University, 525 Dave Lab., University Park, PA 16802, USA}

\emailAdd{malloryr@gmail.com}

\emailAdd{subirb@barc.gov.in}
\affiliation[7]{Astrophysical Sciences Division, Bhabha Atomic Research Centre, Trombay, Mumbai - 400085, India}

\emailAdd{shi.dai@csiro.au}
\affiliation[8]{Australia Telescope National Facility, CSIRO, Space and Astronomy, PO Box 76, Epping, NSW 1710, Australia}

\emailAdd{rene.breton@manchester.ac.uk}
\affiliation[9]{Professor of Astrophysics, University of Manchester, UK}

\emailAdd{ben.stappers@manchester.ac.uk}

\emailAdd{andrea.possenti@inaf.it}
\affiliation[10]{Osservatorio Astronomico Cagliari - INAF, Cagliari, Sardinia, Italy}

\emailAdd{J.W.T.Hessels@uva.nl}
\affiliation[11]{Canada Excellence Research Chair in Transient Astrophysics, McGill University, Montreal, QC, Canada}

\emailAdd{u3009913@connect.hku.hk}

\emailAdd{m.abdelmaguid@nyu.edu}

\abstract{Produced by the interaction between the ``pulsar wind'' powered by the rotational energy of a neutron star and its surroundings, the study of pulsar wind nebulae (PWNe) provides vital insight into the physics of neutron star magnetospheres and ultra-relativistic outflows.  Spatially-resolved studies of the continuum and polarized radio emission of these sources are vital for understanding the production of $e^\pm$ in the magnetospheres of neutron stars, the acceleration of these particles (and potentially baryons) to $\gtrsim10^{15}~{\rm eV}$ energies, and their propagation within the PWN and in the surrounding interstellar medium.  The significant improvements in sensitivity, dynamic range, timing capabilities offered by the Square Kilometer Array have the potential to greatly improve our understanding of the origin of some of the highest energy particles produced in the Milky Way.}

\begin{document}
\maketitle

\section{Introduction}
The notion that neutron stars (NSs) lose the bulk of their rotational energy through a highly relativistic particle outflow generated by their magnetospheres was made even before (e.g., \citealt{pacini68}) the discovery of these objects (e.g., \citealt{hewish68}).  The interaction between this ``pulsar wind'' and the surrounding environment results in a pulsar wind nebula (PWN), a rare class of astronomical objects detected across the entire electromagnetic spectrum. Radio emission from the Crab Nebula (the first entry in Messier's 1774 catalog)  was discovered at a frequency $\nu \sim 100~{\rm MHz}$ \citep{bolton48}, and has since been detected at photon energies as high as $E_\gamma \sim 1~{\rm PeV}$ \citep{lhaaso21} -- a frequency range spanning 22 orders of magnitude.  Soon after it was realized that emission from these objects is primarily non-thermal in nature (e.g., \citealt{shklovsky58}), where lower energy photons ($\lesssim$0.1~GeV or so) are synchrotron radiation from electrons and positrons $e^\pm$ interacting with the nebular magnetic field, while higher photon energies result from these $e^\pm$ inverse Compton scattering lower energy photons which potentially originate from a variety of sources (e.g., the Cosmic Microwave Background, warm dust, and nearby stars, see \citealt{gelfand17} and references therein for a recent review).  In this Chapter, we describe how the improved radio observations of these sources enabled by the SKA will be essential for better understanding the origin of the highest energy photons and particles in the Milky Way and similar star-forming galaxies. 

\section{Current Understanding}
\label{sec:current}
Since the discovery of the Crab Nebula around 250 years ago (e.g., \citealt{messier81}), much has been learned about the nature of PWNe through the meticulous study of the properties of these objects across the electromagnetic spectrum.  In \S\ref{sec:evolution}, we describe our current understanding regarding the evolution of these objects.  In \S\ref{sec:demo}, we describe how this evolution affects the population of PWNe detected in each waveband.

\subsection{Evolutionary Sequence}
\label{sec:evolution}

\begin{figure}[!h]
    \begin{minipage}[t]{0.475\textwidth}
        \vspace{0pt}
        \includegraphics[width=0.975\textwidth]{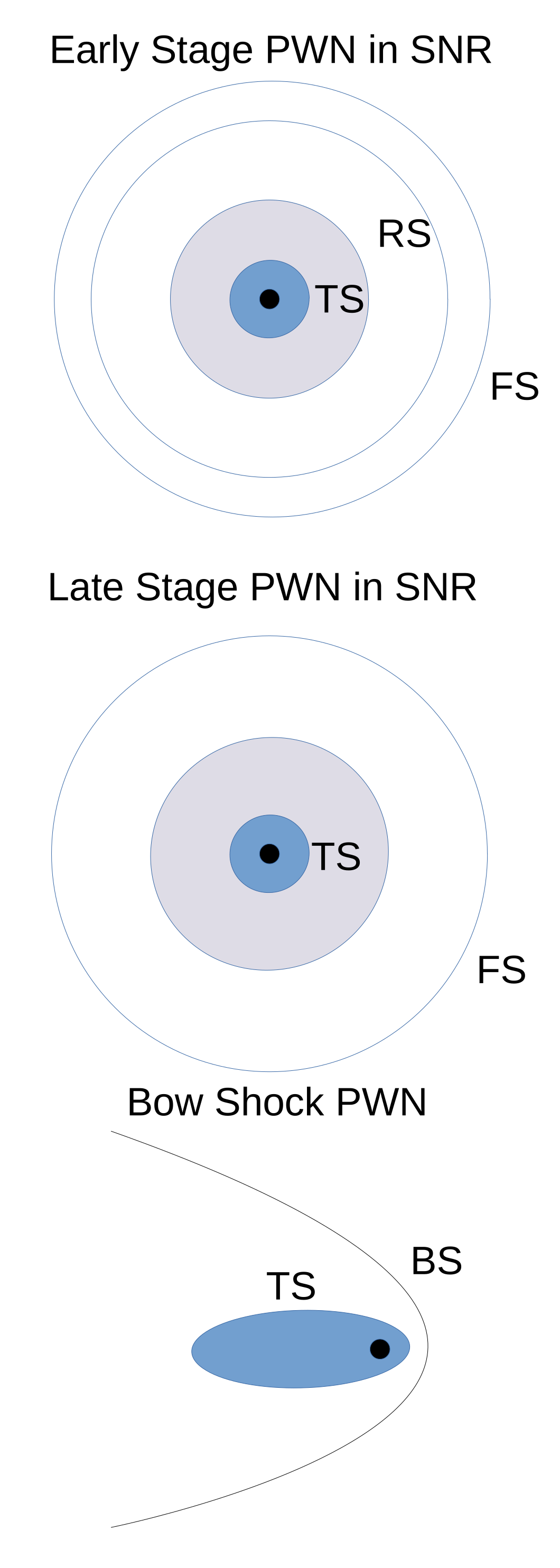}
    \end{minipage}
    \begin{minipage}[t]{0.525\textwidth}
        \vspace{0pt}
        \includegraphics[width=0.85\textwidth]{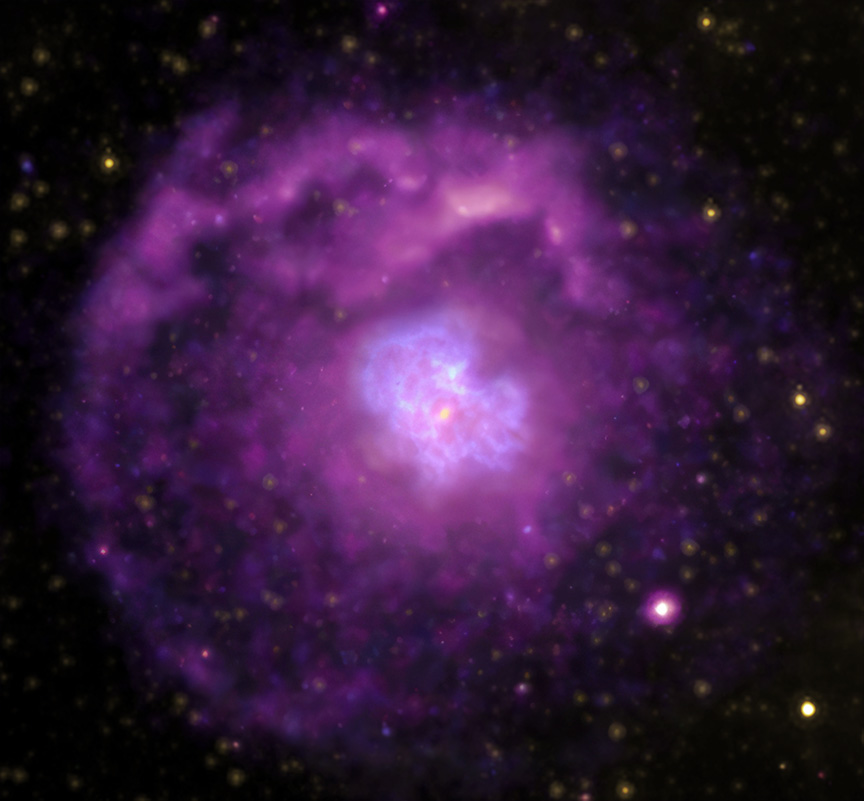}
        \includegraphics[width=0.85\textwidth]{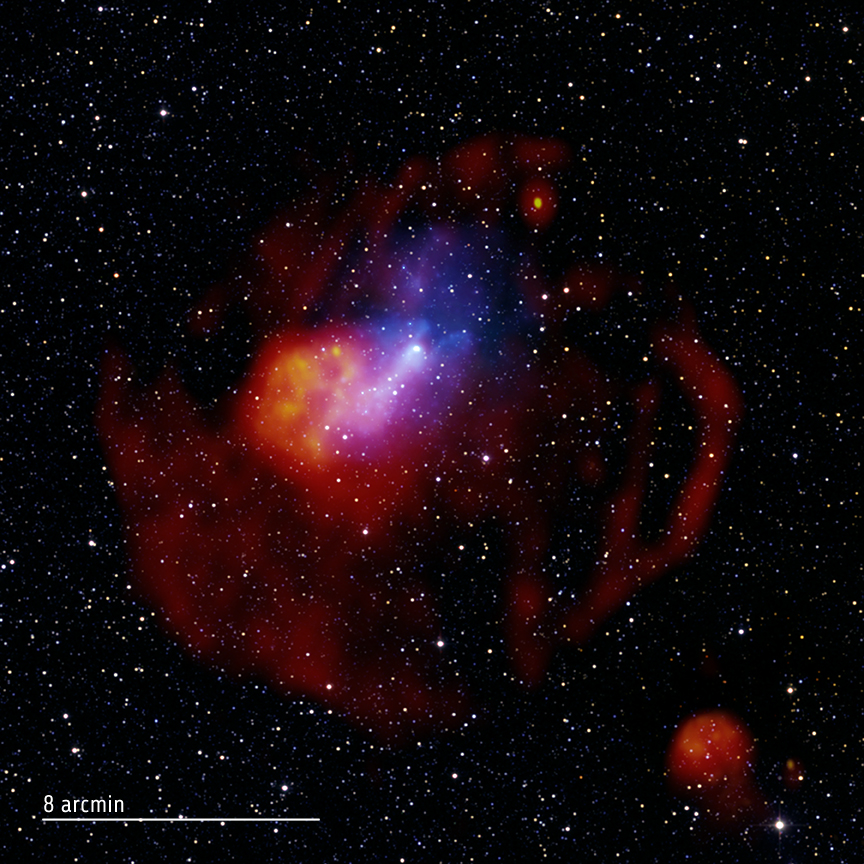}
        \includegraphics[width=0.85\textwidth]{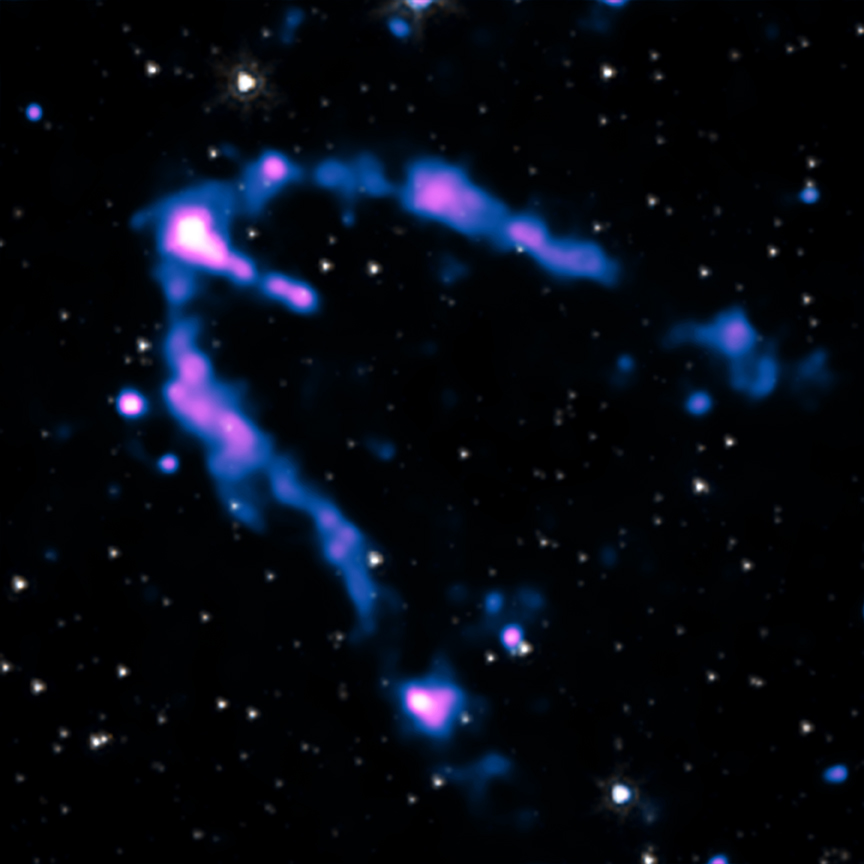}
    \end{minipage}
    \caption{Schematic diagrams ({\it left}) and images ({\it right}) of the PWN generated by an isolated pulsar in the different evolutionary stages.  ({\it Top}): Stage 1, showing the wind termination shock (TS), PWN forward shock, SNR reverse (RS) and forward (FS) shocks.  An example of a PWN in this system is the one associated with SNR G21.5$-$0.9 (Credit: X-ray: NASA/CXC/SAO; Infrared: NASA/JPL/CalTech/Spitzer; Radio: NSF/NRAO/VLA; Image Processing: NASA/CXC/SAO/L. Frattare.). ({\it Middle}): Stage 2, after the SNR reverse shock has reached the PWN, when the SNR has reached the Sedov phase.  An example of a PWN in this stage is the one associated with SNR G327.1-1.1 (Credit: X-ray: NASA/CXC/SAO/T.Temim et al. and ESA/XMM-Newton Radio: SIFA/MOST and CSIRO/ATNF/ATCA; Infrared: UMass/IPAC-Caltech/NASA/NSF/2MASS). ({\it bottom}) Bow shock stage.  An example is the PWN produced by Geminga. (Credit Geminga Images: X-ray: NASA/CXC/PSU/B.Posselt et al.; Infrared: NASA/JPL-Caltech)}
    \label{fig:PWN_stages_schematic}
\end{figure}

The evolution of these systems depends not only on the spin-down evolution of the NSs -- which is the source of energy within the PWN -- but also the particle and magnetic content of its ``pulsar wind,'' the location and morphology of the particle acceleration site, and the nature of its environment (e.g., \citealt{gelfand17} and references therein).  The environment of the PWN is expected to change significantly over its lifetime, as shown in Fig. \ref{fig:PWN_stages_schematic}).  Initially, the PWN is embedded within the unshocked ejecta created by the explosion of the neutron star's stellar progenitor.  Later, the PWN will be surrounded by the hot (primarily ejecta) material generated by passage of the reverse shock (RS) within the supernova remnant (SNR) resulting from the interaction of the ejecta with the surrounding interstellar medium (ISM).  Lastly, the space velocity imparted onto the NS during the supernova explosion causes it to leave the SNR, at which point the pulsar wind directly interacts with the surrounding ISM.  If the neutron star is in a binary system, the evolution of the PWN is further influenced by the interaction between the pulsar wind and its companion, whose nature can vary widely between systems.

As described in a recent review by \citealt{mitchell22}, and numerous references therein, these changes in the surrounding environment result in significant changes in the broadband properties of the PWN.  Initially, when the PWN is surrounded by unshocked ejecta (i.e., a ``Stage 1'' PWN ; e.g., \citealt{mitchell22}), 
the PWN is well described as a bubble of $\sim100\%$ pulsar wind material centered on the associated NS, surrounded by a shell of swept-up ejecta (e.g., \citealt{chevalier05, temim17, temim24}).  At the earliest times, the intense emission generated by the pulsar wind nebula can significantly impact the surrounding supernova ejecta -- especially in cases where the initial rotation energy of the neutron star is comparable to the initial kinetic energy of the ejecta as might be the case for kilonovae (e.g., \citealt{ren19}).

During this stage, the extent of the PWN is typically comparable at all wavelengths though, in cases where the PWN's magnetic field is sufficiently strong, the shorter synchrotron lifetime of the higher energy X-ray emitting particles may already result in a smaller size in this waveband than at radio and $\gamma$-ray energies. 
X-ray observations reveal jet-torus morphologies that represent highly complex plasma distributions in the inner part of young PWNe (e.g., \citealt{Reynolds2017,Olmi2023}). They are manifestations of anisotropic pulsar winds with a non-uniform distribution of the wind energy flux (e.g., \citealt{Komissarov2004}).  

The properties of the PWNe change significantly once it begins interacting with the SNR reverse shock (RS).  Due to the space velocity of the NS, and asymmetric propagation of the RS into the SNR resulting from inhomogeneities in the ISM, regions of the PWN come into contact with RS-heated material at different times.  Since the pressure of the RS-heated material is much higher than the pressure inside the PWN, the expansion of the PWN slows considerably in directions where it has encountered the RS -- resulting in an offset between the NS and the center of the PWN, as reflected in the relative position of the radio and X-ray emisson from sources where this collision is believed to have occurred (e.g., \citealt{gelfand07}).  Once the entire PWN reaches the RS, it begins to contract due to the pressure of the surrounding hot gas -- with the competing adiabatic heating and increased synchrotron losses (resulting from the stronger nebular magnetic field due to this compression) leading to significant changes in broadband spectral energy distribution (SED) of the PWN (e.g., \citealt{gelfand09, bandiera23a}).

As the PWN is compressed, the growth of instabilities along the outer boundary of the PWN results in a considerable change of its composition -- mixing the surrounding material with the pulsar wind injected by the central NS, but also enhancing the rate at which particles leave the PWN and propagate through the surrounding medium -- leading to the eventual disruption of the initial PWN produced by the NS (e.g., \citealt{blondin01, bucciantini04, gelfand09}).  During this stage of their evolution (``Stage 2''; e.g., \citealt{mitchell22}), the X-ray emission is detected beyond the confines of the radio PWN (e.g., \citealt{temim13, temim15}) -- observational evidence of particle escape and interacting with the surrounding (instead of nebular) magnetic field.  The onset of this phase is also believed to trigger the formation of the TeV halos -- extremely large regions of diffuse $\gamma$-rays -- increasingly detected in the ``middle aged'' (characteristic age $\sim10^4-10^5$ year) pulsars (e.g., \citealt{abeysekara17,albert23} , see \citealt{mitchell22} for a recent review).

Eventually, the NS will leave the SNR, and its pulsar wind will directly interact with the surrounding ISM.  In this case, the high energy particles generated by the NS are not always fully confined within a PWN (as when the NS is located within the SNR), but pass through the complex shock structure generated by the NS's motion through its environment (e.g., \citealt{bucciantini18}).  At radio wavelengths, such sources typically have the cometary morphology (e.g., \citealt{gaensler04, camilo09, lazarevic24}) resulting from the supersonic motion of the NS.  Particles injected in the direction of the pulsar's velocity are predominantly confined within the forward bow shock, although a substantial fraction can penetrate through and emit X-rays far in front of the bow shock. Those particles injected antiparallel to the pulsar's velocity may escape even more easily into the surrounding medium.
Evidence for a complicated interaction with the surrounding environment comes from the wide variety of X-ray morphologies observed from fast-moving pulsars, including simple bow shock structures, multiple tails, large-scale extended emission, and long ``filaments'' that are not aligned with the proper motion of the NS (e.g., \citealt{bandiera08, Kargaltsev2017, Klingler2018, devries22}). These morphologies are thought to be due to differences in the pulsar wind itself (wind energy distribution, magnetization), the surrounding medium (its density, magnetic field strength and direction), and geometry of the pulsar's magnetic field, rotation axis, motion vector, and viewing direction which affect the escape of particles into the surrounding medium (e.g., \citealt{Reynolds2017,bucciantini18, barkov19}). 
In addition to X-ray and radio emission, bow shocks have also been detected in H$\alpha$ (see \citealt{Brownsberger2014} and references therein) and more recently at ultraviolet wavelengths (see, e.g., \citealt{Kargaltsev2017}). 

\begin{figure}[tbh]
    \centering
    \includegraphics[width=0.75\textwidth]{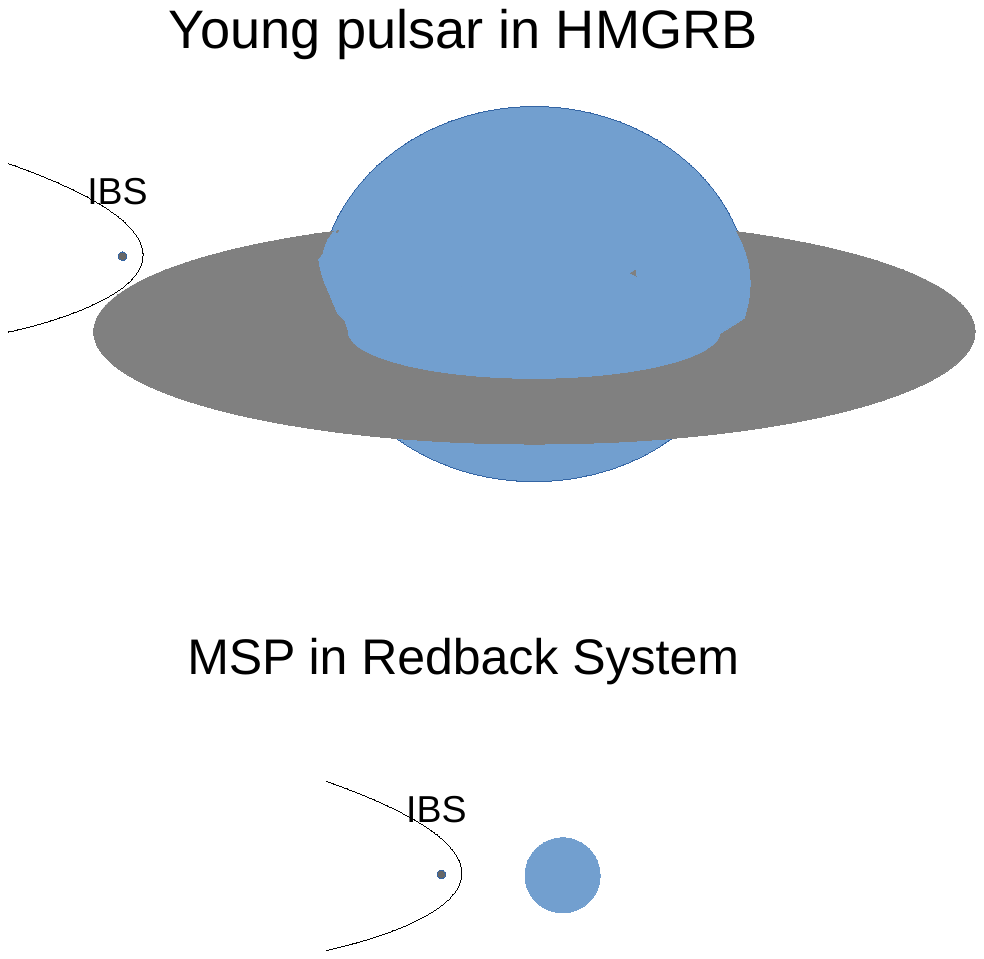}
    \caption{Intrabinary pulsar wind shock (IBS) examples: ({\it top}) a High Mass Gamma-Ray Binary where a young pulsar interacts with the excretion disk of a massive companion; the orbit will likely be highly eccentric; ({\it bottom}), a redback system where a recycled MSP is in a close, circular orbit around a non-degenerate low mass companion.}
    \label{fig:binaries}
\end{figure}

For NSs in binary systems, the companion provides an alternative environment to the ISM for pulsar wind interactions, as well as a different evolution for the neutron star (Fig.\ref{fig:binaries}). If the companion is a massive star (see e.g., \citealt{chernya2019} and references therein), right after the SN explosion, and if the binary is not disrupted, the neutron star will be in a highly eccentric orbit in an environment dominated by the companion's outflows. This can be in the form of a strong wind or, in the case of Oe and Be stars, an equatorial excretion disk. The young pulsar will have a strong wind, but may or may not be visible as a radio pulsar due to excess scattering material in the system. At first, the pulsar's wind, strong magnetic field, and fast spin will prevent material from accreting onto the neutron star, but eventually the pulsar will slow, the wind decrease, and accretion will occur, extinguishing the wind and radio pulsations but generating X-ray pulsations, becoming a classic X-ray Binary (XRB) Pulsar.  At first, material from the companion slows down the neutron star rotation even more, but as it overcomes magnetic inhibition it will spin up the pulsar until the companion undergoes its own SN explosion, the end result possibly being a NS-NS binary containing a partially recycled pulsar ($P_s \sim 10-50$~ms). 

If the companion is an intermediate or low mass star, the evolution is similar, but the accretion episode lasts for much longer, resulting in the neutron star being spun up to millisecond periods and ending with a NS in a highly circular orbit around a white dwarf where the neutron star spin axis is aligned with the orbital axis (see e.g., \citealt{lorimer08} and references therein). This accretion is not steady, however, and the system can detach intermittently before the companion has reached the white dwarf stage. During these detached times, radio pulses and the pulsar wind can turn back on. After the companion becomes a white dwarf, the neutron star is now a recycled pulsar, generally with a spin period of a few milliseconds (a millisecond pulsar, MSP) and emitting radio through $\gamma$-ray pulsations. If the companion is close enough, the pulsar will begin to ablate the companion, potentially eventually completely evaporating and leaving behind an isolated MSP.

In a binary system, the pulsar wind can interact with both the ISM and its companion to produce observable broadband emission. In general, radio pulsations are eclipsed for part of the orbit, and emission from an intrabinary shock is  strongly orbitally modulated.  This is the case for both young pulsars around massive companions and recycled (millisecond) pulsars around low-mass companions (see \citealt{rat+18} and references therein). The young systems are an important subclass of High Mass Gamma-ray Binaries (HMGB) which, in principle, may have either neutron stars or black holes orbiting a massive companion. In the HMGB that contain a pulsar, the pulsar wind interacts with a strong companion wind. The companion also provides a bright source of photons that can be inverse Compton scattered to $\gamma-$ray energies by the pulsar wind. However, there are only a few cases where radio pulses have been seen and so the nature of the compact object is  definitively known. In some cases, VLBI imaging have revealed that there is compact PWN emission in addition to emission from an intrabinary shock.  The recycled systems with intrabinary shock emission are the subclass of binary millisecond pulsars known as spiders, which are systems where the pulsar wind is ablating the companion, generating excess material both within and outside the binary orbit.  In principle, there are multiple advantages to studying pulsar winds in binary systems. First off, the interaction can be observed from multiple angles within one system and is repeated every orbit. Perhaps more importantly, shocks are forced to occur at  distances which probe the pulsar wind much closer to the light cylinder than is the case in PWN around isolated pulsars, and by studying the population, you can probe different radii in terms of the fundamental parameter of the light cylinder radius. In the case of the HMGB, you can even probe a range of distances within one system since the pulsar orbit tends to be highly eccentric.  

Of particular interest and value are double neutron star systems that are seen nearly edge on. The passage of radio pulses from the partially recycled pulsar through the magnetosphere of its companion provides crucial information on the spatial distribution of charged particles in this region and the shape of the radio beam (e.g., \citealt{lyutikov05, rafikov05, yi17, lower24}).

\subsection{PWN demographics}
\label{sec:demo}
As mentioned in \S\ref{sec:evolution}, the SED of a PWNe changes considerably over the course of its evolution.  Therefore, the properties of PWNe currently detected differ between wavebands as described below.  

The radio and X-ray emission from PWNe is believed to be dominated by synchrotron radiation from $e^{\pm}$ interacting with nebular magnetic field.  The characteristic frequency $\nu_c$ of such emission from a particular particle depends on both its energy $E$ and the magnetic field strength $B$ at its location, with:
\begin{eqnarray}
    \label{eqn:nu_c}
    \nu_{\rm c} & \approx & 10.7 \left(\frac{B}{\rm 1~\mu G} \right) \left(\frac{E}{\rm 1~GeV}\right)^2~{\rm MHz},
\end{eqnarray}
and these particles will lose most of their energy as synchrotron radiation in time:
\begin{eqnarray}
    \label{eqn:tsynch}
    t_{\rm synch} & \approx & 8\left(\frac{B}{\rm 1~\mu G} \right)^{-2} \left(\frac{E}{\rm 1~GeV} \right)^{-1} \times 10^{9}~{\rm years},
\end{eqnarray}
commonly referred to as the synchrotron lifetime of these particles.  Combined, Equations \ref{eqn:nu_c} \& \ref{eqn:tsynch} imply that the radio synchrotron emission from a PWN will originate from $e^\pm$ with a lower energy $E$ and longer lifetime $t_{\rm synch}$ than the $e^\pm$ responsible for the X-ray synchrotron emission of such sources.


The maximum synchrotron photon energy can be expressed as
\begin{eqnarray}
\label{equ:emax}
E_{\rm max} \lesssim 130 \zeta \dot{E}_{35} B_{-5} \sigma/(\sigma +1) ~{\rm keV}
\end{eqnarray}
(e.g., \citealt{Reynolds2017}), where $\zeta \sim 1$ is a numerical factor, $\dot{E}_{35}$ is the spin down power of the pulsar in units of $10^{35}$\,erg/s, $B_{-5}$ is the magnetic field in $10^{-5}$\,G, and $\sigma$ is the poorly known magnetization parameter, the ratio of the Poynting flux to the particle enthalpy flux. Characteristic PWN magnetic field values are thought to be roughly in the range of $1 - 100$\,$\mu$G and are typically higher in shocked PW regions. From the dependency on the spin-down energy in equation~\ref{equ:emax}, one expects bright synchrotron X-ray PWNe only around the youngest pulsars, with radio and IC $\gamma$-ray emission lasting for much longer.  Around even older (and very close) pulsars one may still detect UV emission and H${\alpha}$ PWNe.

The PWN's magnetic field can be strong, $\mathcal{O}(100~\mu~{\rm G})$, as inferred for some PWNe associated with young ($\lesssim 10^3$~years) pulsars as well as for the PWNe in some binary systems. For such PWNe, their X-ray synchrotron emission ($h\nu \sim 1-10~{\rm keV}$, $\nu \sim (3-30)\times10^{17}$) is emitted by $e^\pm$ with energy $E\sim 10-100~{\rm TeV}$.  For PWNe with weaker magnetic field, e.g. $\mathcal{O}(\rm \mu G)$, their X-ray synchrotron emission is produced by particles with energy $E\sim 0.1 - 1~{\rm PeV}$.  The $\sim10^{4}-10^{5}$~yr synchrotron lifetime of these particles is comparable to, or longer than, the $\sim10^3 - 10^5$~yr characteristic age of the neutron stars often associated with PWNe, suggesting that X-ray emission could be produced throughout the PWN.  However, the relatively small number of $e^\pm$ with the required extreme energies can result in a surface brightness lower than the sensitivity of existing observations and facilities.  


The situation is very different for the PWN's synchrotron emission at the frequencies covered by SKA-Low ($\nu \sim 50-350~{\rm MHz}$) and SKA-Mid ($\nu \sim 0.35-15.4~{\rm GHz}$).  For strong-field PWNe, this frequency range is produced by $e^{\pm}$ with energy $E \sim 0.2-1~{\rm GeV}$, which have a synchrotron cooling time of $t_{\rm synch} \sim 10^6$~yr -- significantly longer than the typical characteristic ages of the isolated neutron stars associated with such objects.  For weaker nebular magnetic field strengths, the radio emission originates from $E \sim 2-20~{\rm GeV}$ $e^\pm$ with even longer synchrotron cooling times ($t_{\rm synch} \sim 10^8-10^9$~yr).  As a result, the PWN's radio emission includes contributions from particles generated throughout its lifetime -- unlike its X-ray emission, which is dominated by more energetic, recently injected particles. In addition to the radiation losses, evolutionary effects of the pulsar may contribute to the observed sharp steepening of the PWN synchrotron spectrum between radio and X-rays.


However, PWNe do not only emit synchrotron radiation.  The $\gamma$-ray emission from PWNe is believed to the result of high energy $e^{\pm}$ inverse Compton (IC) scattering off background photons originating from the Cosmic Microwave Background (CMB) as well as emission from nearby dust (far- and mid-infrared) and stars (predominantly near-infrared and optical).  If the CMB is the dominant source of target photons, then a $\gamma$-ray of energy $E_\gamma$ is produced by scattering off an $e^\pm$ with energy:
\begin{eqnarray}
    \label{eqn:egammae}
    E & \approx & 63 \sqrt{\frac{E_\gamma}{\rm 10~TeV}}~{\rm TeV}
\end{eqnarray}
which, if it only lost energy due to IC emission, has a radiative lifetime of:
\begin{eqnarray}
    \label{eqn:tic}
    t_{\rm ic} & \approx & 30 \left(\frac{E}{40~{\rm TeV}} \right)^{-1} \times 10^3~{\rm years}. 
\end{eqnarray}
This suggests that the $\sim 0.1 - 10~{\rm TeV}$ emission detected from a large and, thanks to new $\gamma$-ray facilities such as HAWC and LHAASO and future ones like the Cerenkov Telescope Array (CTA), growing number of Galactic PWNe originates from $E \sim 20-60~{\rm TeV}$ particles with an IC lifetime of $t_{\rm ic} \sim  10^{4.5} - 10^6$~years.

The particle energies and lifetimes estimated above are useful in understanding how the multi-wavelength properties of PWNe change as they progress through the evolutionary sequence described in \S\ref{sec:evolution}.  Soon after the core-collapse supernova that creates the neutron star, the strong magnetic field of its PWNe will result in X-ray emitting particles having a radiative lifetime comparable to, or shorter than, the age of the system.  Due to the transport of particles away from the termination shock where they are injected into the PWN, X-ray emission detected further from this structure is radiated by particles injected earlier.  This fact, and the rapid cooling of these $\sim10-100$\,TeV e$^\pm$, results in a relative deficit of high energy e$^\pm$ at larger distances from the termination shock -- and explains in large part the observed correlation between the X-ray luminosity of a PWN ($L_X$) and the spin-down luminosity $\dot{E}$ of its associated neutron star \citep{Kargaltsev2008b}. Such cooling explains the decrease in X-ray surface brightness $\Sigma_X$ and ``softening'' of the X-ray spectrum (increase in photon index $\Gamma$) observed in such systems.  As described below, these measurements can be used to test and constrain models for the transport of particles within the PWN (e.g., \citealt{kundu24}).  For these strong magnetic fields, the particles energetic enough to produce X-ray synchrotron emission (Equation \ref{eqn:nu_c}) are also sufficiently energetic to produce $\gamma$-ray inverse Compton emission (Equation \ref{eqn:egammae}, which suggests that similar spatial variations in the surface brightness and spectrum should be observed at X-ray and $\gamma$-ray energies, though they are hard to measure observationally due to the large point spread function of current facilities.  However, the long lifetime of the radio-emitting particles suggests emission in this waveband should originate from the entire PWN, and therefore can have an angular extent larger than detected at higher photon energies. However, empirically the TeV emission is often observed to have the largest extent (see below). 

For PWNe with weaker magnetic fields, both the X-ray and $\gamma$-ray emission originate from $\gtrsim100~{\rm TeV}$ particles.  Since IC scattering is the dominant emission mechanism of such particles, such PWN are likely to have a higher surface brightness at $\gamma$-ray energy $\Sigma_\gamma$ than at X-ray energies, though emission at both wavebands will be hampered by the rarity of $e^\pm$ with the required high energies.  Since the angular size of such PWNe are often larger than the field of view of current X-ray telescopes but smaller than that of $\gamma$-ray facilities, their $\gamma$-ray properties are often better measured than their X-ray properties.  Radio emission should also be generated throughout such PWNe.  However, it can be difficult to detect due to the low surface brightness of this emission being hard to distinguish against the bright Galactic plane background. Also,  the angular size of PWNe may be larger than the largest angular scale of the radio interferometers used in such studies.

For NSs directly interacting with the ISM, their anisotropic pulsar wind emission,  typically faster-than-sound motion, as well as inhomogeneities in the ambient medium and ISM entrainment are known to produce striking examples of detected X-ray head-tail structures as well as multiple tails and PWN shapes around pulsars that have left their parent supernova remnant (e.g., \citealt{Kargaltsev2017}). \citet{Kargaltsev2008} found that the X-ray efficiencies of the ram pressure-confined PWNe are systematically higher than those of PWNe around slowly moving pulsars with similar spin-down parameters.   Nevertheless, pulsars with low spin-down power are not expected to harbor strong X-ray PWNe. When such PWN are seen in radio, the radio emission is generally better defined by the bow-shock than the X-ray emission.


%
%
%




\section{Open Questions}
\label{sec:open}

While the notion that PWNe are primarily powered by the rotation energy of neutron stars was established nearly $\sim6$ decades ago (e.g., \citealt{pacini68, goldreich69}), as described below their remains several unanswered fundamental questions regarding the nature of these sources, especially concerning their ability to generate the highest energy photons -- and therefore, among the highest energy particles -- observed from the Milky Way.

\subsection{Generation of the pulsar wind}
\label{sec:wind}

Currently, it is believed that both the ``pulsar wind'' and pulsed emission are generated by particles produced in the neutron star magnetosphere.  A neutron star's combination of a rapidly rotating surface and a strong magnetic field generates a voltage strong enough to remove charged particles from the surface (e.g., \citealt{goldreich69}, see also a
review by \citealt{lorimer12}).   These particles produce photons as they travel along curved magnetic fields above the neutron star surface (often referred to as ``curvature radiation;'' e.g., \citealt{ruderman75}).  These photons not only contribute to the pulsed radio emission generated by a neutron star (e.g., \citealt{gil04}), their interaction with the strong magnetic fields results in their conversion to an $e^\pm$ pair.  These pairs also generate curvature radiation which, if produced in a region with sufficiently strong magnetic field, also pair produce.  As a result, near the surface the neutron star magnetosphere is filled with a dense, primarily $e^\pm$, plasma (e.g., \citealt{goldreich69}) whose motion is dictated by the structure of the neutron star's magnetic field.  It is those particles moving away from the neutron star surface along ``open'' field lines which comprise the pulsar wind whose interaction with the surrounding medium generates the PWN.  Therefore, the particle content of a PWN directly probes the pair production mechanism inside the neutron star magnetosphere.

The number of particles inside a PWN provides information on the ``multiplicity'' -- i.e., the efficiency of the pair production process described above.  Current theoretical models suggest the number of particles generated in the magnetosphere depends on a number of physical parameters (e.g., \citealt{timokhin15, timokhin19}), including the magnetic field strength -- which can be estimated from the timing properties of the neutron star (see \citealt{lorimer12}), or measured from cyclotron lines in X-rays (e.g., \citealt{Staubert2019}), the curvature of the magnetic field lines near the surface, and the temperature of the neutron star surface -- which can be measured from X-ray observations (see a recent review by \citealt{rigoselli25}).    

Measuring the number of $e^\pm$ inside a PWN requires analyzing its SED.  However, the amount of synchrotron emission generated by a PWN is degenerate with particle number and nebular magnetic field strength, while its inverse Compton luminosity is degenerate with particle number and the total energy density of the target photon fields -- which in some cases can greatly exceed that of the CMB (e.g., \citealt{straal23}).  As a result of these degeneracies, multiplicities estimated using emission in only one waveband depend on a large number of assumptions (e.g., \citealt{spencer25}), and robust estimates of the multiplicity require studying the broadband SED.  The radio emission from PWNe is particularly important for such work, since the long radiative lifetime of these particles results in the total radio emission being sensitive to the number of particles generated over its entire history, and the lower energy of radio-emitting electrons suggests they dominate the total number of particles in the PWN (e.g., \citealt{gelfand15}).

Studies of PWNe can also determine the ionic composition of the pulsar wind.  While the particle production mechanism described above produces $e^\pm$, the electric potential generated in the magnetosphere is sufficient to also extract ions from neutron star surface.  Therefore, they might also be an ionic component to the pulsar wind (e.g.,  \citealt{ostriker69,arons98}) -- though observations suggest the ions must be energetically subdominant to the $e^\pm$ in these sources (e.g., \citealt{spencer25}).  Direct evidence for ions inside PWNe would be the detection of high-energy neutrinos from these sources -- potentially possible with the next generation of such facilities (e.g., \citealt{amato03}).  However, the structure of, and spectrum of particles, within PWNe might also offer indirect evidence for ions in the pulsar wind (e.g., \citealt{arons98}).  The motion of ions within the PWN has been hypothesized to result in the variable ``wisps'' observed in these objects primarily at optical (e.g., \citealt{melatos05}) and X-ray wavelengths (e.g., \citealt{gallant94, spitkovsky04}).  In addition, since ions in the pulsar wind will also impact the acceleration of particles in the nebula (e.g., \citealt{amato06}), possibly creating signatures in the energy spectrum of $e^\pm$ in such sources (e.g., \citealt{guepin20}).


\subsection{Acceleration of particles inside the PWN}
\label{sec:acc}

The increasing number of associations between Galactic PeVatrons -- sources which emit $>100~{\rm TeV}$ photons -- with PWNe (e.g., \citep{ohira18, desarkar22, park23, woo23}) requires that particles are accelerated to 
$\gtrsim$PeV energies (e.g., \citealt{cao21, wilhelmi24}) in these objects.  Currently, it is believed that the primary site of particle acceleration inside a PWN is the ``termination shock'' formed where the ram pressure of the unshocked pulsar wind is equal to that of the surrounding PWN (e.g., \citealt{rees74, kennel84} and citations thereafter) -- with the resultant particle spectrum believed to be sensitive to the properties of the unshocked pulsar wind at this location (e.g., \citealt{sironi09, sironi11}).

When the pulsar wind leaves the neutron star magnetosphere, it is expected to be magnetically dominated and launched at high rotational latitudes. The magnetic field outside the light cylinder is forced to become toroidal and concentrated along the spin equator which channels the wind into an equatorial outflow.  Since the existence of pulsations requires that the magnetic axis is tilted with respect to the spin axis, 
the pulsar wind is ``magnetically striped'' (i.e., the direction of its magnetic field regularly changes direction by $180^\circ$;   
) 
Inside the PWN, past the termination shock,  the observed X-ray spectra and emission energetics suggest that the particles are energetically dominant (e.g., \citealt{kennel84, bucciantini11, torres14, abdelmaguid23}). 
This transition from a magnetically to kinetically dominant outflow is thought to be intimately related to the acceleration of particles in these sources.  Most models postulate that the dissipation of the striped magnetic wind  happens upstream (i.e., between the magnetosphere and termination shock, e.g., \citealt{coroniti90, kirk03, cp17, cpd20}), but others suggest that it may occur at the termination shock itself (e.g., \citealt{petri07}) or within the pulsar wind nebula itself (e.g., \citealt{porth13, zrake17}). In these latter cases, the wind may remain magnetically striped up to the termination shock. If correct, this suggests that particles may be (re-)accelerated throughout the nebula (e.g., \citealt{lyutikov19}). 

Observations suggest the acceleration mechanism within a PWN produces particles with a ``broken power law'' energy spectrum $\frac{dN}{dE} \propto E^{-p}$ where $p$ has different values above and below some characteristic break energy $E_{\rm b}$ (e.g., \citealt{bucciantini11, torres14, gelfand15}) -- with the particle index $p_1$ at energies below the break energy $E_{\rm b}$ typically (e.g., \citealt{bucciantini11, torres14, gelfand15}), but not always (e.g., \citealt{hattori20}), lower than the paricle index $p_2$ at high energies -- i.e., $p_1 < p_2$.   Such a particle spectrum is currently thought to arise from a  combination of ``standard'' Fermi acceleration mechanism and magnetic reconnection (e.g., \citealt{sironi11, cerutti20}) within the PWN -- with magnetic reconnection responsible for the ``hard'' particle spectrum at low energies inferred from the radio observations of these sources (e.g., \citealt{lyutikov19}).

While theoretical work suggests that magnetic reconnection is capable of producing the ``hard'' particle spectra observed in PWNe (e.g., \citealt{bessho12, werner21, french23, zhang23}), the acceleration efficiency and particle spectrum generated by this mechanism depends strongly on the properties on the particle flow (e.g., \citealt{cerutti14, sironi14, lu21}).  In particular, turbulence -- especially in the kinetic limit -- is found to inhibit magnetic reconnection \citep{LB2018APJ}, which would have a significant impact on the spectrum of particles inside a PWN.


Understanding the acceleration mechanism requires measuring the particle spectra in a sample of PWNe with a large range in pulsar wind properties.   In binary systems, the pulsar wind environment is determined primarily by the companion's wind, magnetic field, and photon emission (the companion is an important source of photons for inverse Compton emission). In the direction of the companion, a shock front is forced to occur within the binary, at a distance  (as measured in light cylinder radii $R_{lc}$) much closer to the pulsar than is the case for PWN around isolated neutron stars ($\sim 10^3 - 10^5\, R_{lc}$ as opposed to $10^8 - 10^9\, R_{lc}$ for PWN, see e.g. \citep{rmg+14,cs22}). For young pulsars around massive companions, the orbital plane of the pulsar is generally not aligned with the equators of either the pulsar or the companion, and the orbit may be highly eccentric. Therefore, the shock occurs over a range of distances and pulsar wind latitudes within one system. For recycled systems, it is probable that the spin axes and the orbital axis are aligned, and the orbit is highly circular. 
Numerical models of the magnetic dissipation radius of the striped wind  suggest that it is similar to the intrabinary shock distance in binary systems \citep{cpd20}. The pulsar wind in binary systems is therefore much more strongly magnetized when it reaches the termination shock than for PWN around isolated pulsars.  This difference, and the resultant impact on the underlying acceleration mechanism, has been invoked to explain the extremely hard particle spectra implied by the X-ray spectra (photon index $\Gamma \sim 1.0-1.4$) of such sources \citep{rat+18}.

\subsection{Transport of particles within and beyond the PWN}
\label{sec:transport}
As mentioned in \S\ref{sec:acc}, the particles generated in the neutron star magnetosphere enter the PWN at the termination shock.  The propagation of these particles through, and eventually beyond, the PWN has implications for the possible re-acceleration of particles inside the nebula as mentioned above, as well as the PWN contribution to the Galactic population of cosmic ray $e^\pm$.  

Initially, it was assumed that (e.g., \citealt{kennel84}) PWN consisted of a MHD flow of particles away from the termination, but these models were found to be inconsistent with observed spatial variations in the X-ray surface brightness $\Sigma_X$ and photon index $\Gamma$ in these sources (e.g., \citealt{reynolds09, tang12}).  As a result, more recent models tested whether a combination of diffusion (e.g., \citealt{porth16, zhu23}) and advection (e.g., \citealt{collins24}) dominate the motion of particles through the PWN.  While such models often can reproduce either the observed $\Sigma_X$ or $\Gamma$ profile within a PWN (e.g., \citealt{kim20}), they also often are unable to reproduce both observed quantities with the same set of model parameters (e.g., \citealt{kundu24}).  This has led to further studies which found that turbulence could play an important role in both distributing and accelerating particles throughout the PWN -- especially at the lower particles energies responsible for the PWN's radio emission (e.g., \citealt{luo20, lu23, lu24}).




Since $\approx100\%$ of the material inside a PWN is electrically charged, the movement of particles will be strongly coupled to the structure of the nebular magnetic field.  The detection of polarized emission (see \citealt{volpi09} for a review) from PWN at radio (e.g., \citealt{kothes20, ma16}), optical (e.g., \citealt{moran14}), and increasingly at X-ray energies (e.g., \citealt{xie22, liu23, xie24, bucciantini25}) suggests a significantly ordered magnetic field.

Radio polarization measurements provide a powerful probe of the nebular magnetic field structure.  Previous observations found a large diversity in the $B$-field geometry. Many young sources exhibit toroidal $B$-field near the inner PWN, including the Crab \citep{velusamy85,aumont10}, G54.1+0.3 \citep{lang10}, 3C~58 \citep{bucciantini25}, and MSH 15-52 \citep{zhang25}. In the first two cases, the $B$-field orientation switches to radial in the outer PWN near the edge \citep{bietenholz91}, while it becomes highly elongated for 3C~58.  In contrast, a purely radial field is found in radio observation of the young PWN G21.5$-$0.9 \citep{lai22}.  On the small scale, strongly polarized synchrotron filaments and loops are often observed in PWNe, and their $B$-field geometry well follows the emission structure \citep[e.g.,][]{lang10,bucciantini25}. These magnetic filaments and loops could result from instabilities which mix the pulsar wind and the surrounding medium \citep{jun96a,jun96b}, or could be torn by kink instabilities from the wind termination shock \citep{slane04}.

For older PWN systems, a toroidal field is often observed, e.g., Vela \citep{dodson03}, the Boomerang \citep{kothes06}, the Dragonfly \citep{jin23}, and the PWN powered by PSR B1706$-$44 \citep{liu23b}.  Intriguingly, the Snail, which is a middle-aged PWN crushed by the supernova reverse shock, shows a highly ordered $B$-field structure with loops and filaments \citep{ma16}.  Finally, various $B$-field configuration is also observed in bow-shock PWNe, although they are governed by relatively simple boundary conditions. Some are found to have $B$-field well aligned with the nebular elongation, e.g., the Mouse \citep{yusef-zadeh05}, the Frying Pan \citep{ng12}, and possible
the Potoroo \citep{lazarevic24}. On the other hand, both G319.9$-$0.7 and G283.1$-$0.59 show helical fields trailing the pulsar \citep{2010ApJ...712..596N,ng17}.

The origin of such diverse $B$-field geometry in PWNe is unclear. It could depend on many physical parameters, including the evolutionary state, flow structure, pulsar orientation. It is critical to expand the sample for further study. The magnetic field structure could have important implications on particle transport and hence the efficiency of cosmic ray production.

Recent $\gamma$-ray observations discovered large scale halos around pulsars, and in many cases, the TeV emission peak shows significant offset from the associated pulsar \citep{albert20,cao24}. This challenges the standard PWN model and requires some mechanisms to transport the freshly accelerated particles to the gamma-ray emitting region, or some re-acceleration process at the TeV site.  Magnetic fields could play an important role in resolving this issue. For bow shocks, the formation of a large-scale, ordered magnetic field aligned with the pulsar tails is expected, as simulations show that the ISM $B$-field can be swept up to create magnetotails, especially for highly magnetized pulsar wind \citep{olmi19}. In such cases, post-shock particles can gyrate along the magnetic field without much diffusion and advection, therefore moving predominantly along the field lines. This creates a magnetic flux tube structure that enables fast transport of particles and it can shape the overall PWN morphology. As the particles are moving along the field lines, their pitch angle distribution is anisotropic. The synchrotron emissivity will be highly beamed and dependent on the viewing geometry \citep{bao24a,bao24b,krumholz24}. Polarimetric measurements with the SKA will provide a powerful probe
of the nebular magnetic field structure to confirm this model.




The structure of the PWN's magnetic field also plays an important role in the escape of particles from the nebula.  The interaction between the PWN's magnetic field and that of its environment is thought to govern the escape of particles from the PWN (e.g., \citealt{bucciantini18, barkov19, toropina19}) -- with reconnection between the two fields providing a channel which enhances particle escape (e.g., \citealt{olmi24}).  This mechanism is believed responsible for the growing number of X-ray filaments observed around neutron stars (see \citealt{dinsmore24} for a recent catalog), which are often misaligned with the neutron stars' proper motion and magnetic axis (e.g., \citealt{bandiera08}) but possibly aligned with the interstellar magnetic field (e.g., \citealt{dinsmore25}).  The recent association between a candidate radio pulsar / X-ray PWN and a non-thermal radio filament \citep{yusef24} suggests a similar origin for many of the numerous radio filaments found near the Galactic Center (e.g., \citealt{heywood22}).

The structure of the interstellar magnetic field is also believed to play an important role in the formation and evolution of ``TeV halos'' --  extremely large regions of TeV $\gamma$-ray emission (e.g., \citealt{abeysekara17}) -- which are expected to be present around a large fraction of neutron stars (e.g., \citealt{sudoh19, albert25}), and believed to be represent the very late stages of PWN evolution (e.g., \citealt{giacinti20}). While the high energy $e^\pm$ responsible for the observed extended $\gamma$-rays are believed to have escaped from the PWN and are now diffusing through the ISM, the diffusion coefficient inferred from the angular distribution of their high-energy emission is consistently found to be significantly lower than the Galactic average (i.e., ``slow diffusion;'' e.g., \citealt{dimauro20, bao22,  schroer23, albert24}).  Since these particles are charged, their motion is again greatly impacted by the structure of the local magnetic field.  The presence of MHD turbulence -- possibly generated by SNR formed in the progenitor explosion (e.g., \citealt{fang19}) -- around neutron stars will significantly impact the propagation of cosmic rays (e.g., \citealt{farmer04, lopezcoto18, maiti22, xu13}), likely significantly lowering the local diffusion coefficient (e.g., \cite{liu19, gao25}).  Currently, the best way of detecting -- and measuring the power spectrum -- of magnetic turbulence in these regions requires studying the polarized synchrotron emission of these particles (e.g., \citealt{LP2016ApJ, LP2012ApJ, pavaskar24})


\section{Opportunities with the SKA telescopes}
\label{sec:ska}
Measuring the radio properties of PWNe is critical for answering the open science questions detailed in \S\ref{sec:open}.  The number dominance of radio-emitting particles makes the total radio emission from a PWN most sensitive to the multiplicity of particles in magnetosphere (\S\ref{sec:wind}) -- complimentary to the magnetospheric studies described by \citet{Oswald01.2026.SKA}.  Furthermore, it is the spectrum of the radio emitting particles which are thought to be most affected by magnetic reconnection and turbulent re-acceleration inside the PWN (\S\ref{sec:acc}).  Lastly, the long radiative lifetime of the radio emitting particles makes emission in this band extremely useful for understanding how particles accelerated in PWNe propagate within the ISM -- which depends strongly on the structure of the Galaxy's magnetic field \citep{Sun01.2026.SKA}, especially on small scales \citep{Ma01.2026.SKA} around the PWN. 


\begin{table}[tb]
    \centering
    \begin{tabular}{cc}
    \hline
    \hline
    {\sc Parameter} & {\sc Value} \\
    \hline
    Initial kinetic energy of supernova ejecta  $E_{\rm sn}$ & $10^{51}~{\rm ergs}$  \\
    Mass of supernova ejecta $M_{\rm ej}$ & $8~{\rm M}_\odot$ \\
    ISM Density $n_{\rm ism}$ & $0.1~{\rm cm}^{-3}$ \\
    Initial Spin-down Luminosity $\dot{E}_0$ & $10^{40}~\rm{ergs~s^{-1}}$ \\
    Spin-down Timescale $\tau_{\rm sd}$ & 500~years \\
    Wind Magnetization $\eta_{\rm B}$ & $10^{-3}$ \\
    Minimum Injected Particle Energy $E_{\rm min}$ & 1 GeV \\
    Break Injected Particle Energy $E_{\rm break}$ & 1 TeV \\
    Maximum Injected Particle Energy $E_{\rm max}$ & 1 PeV \\
    Low Energy Particle Spectrum $p_1$ & 1.5 \\
    High Energy Particle Spectrum $p_2$ & 2.5 \\
    \hline
    \hline
    \end{tabular}
    \caption{Input parameters to the \citet{gelfand09} model for the evolution of a PWN inside a SNR used to predict the properties of the PWN's radio emission shown in Figure \ref{fig:radio_evol}.}
    \label{tab:model_parameters}
\end{table}

\begin{figure}[tbh]
    \centering
    \includegraphics[width=0.475\textwidth]{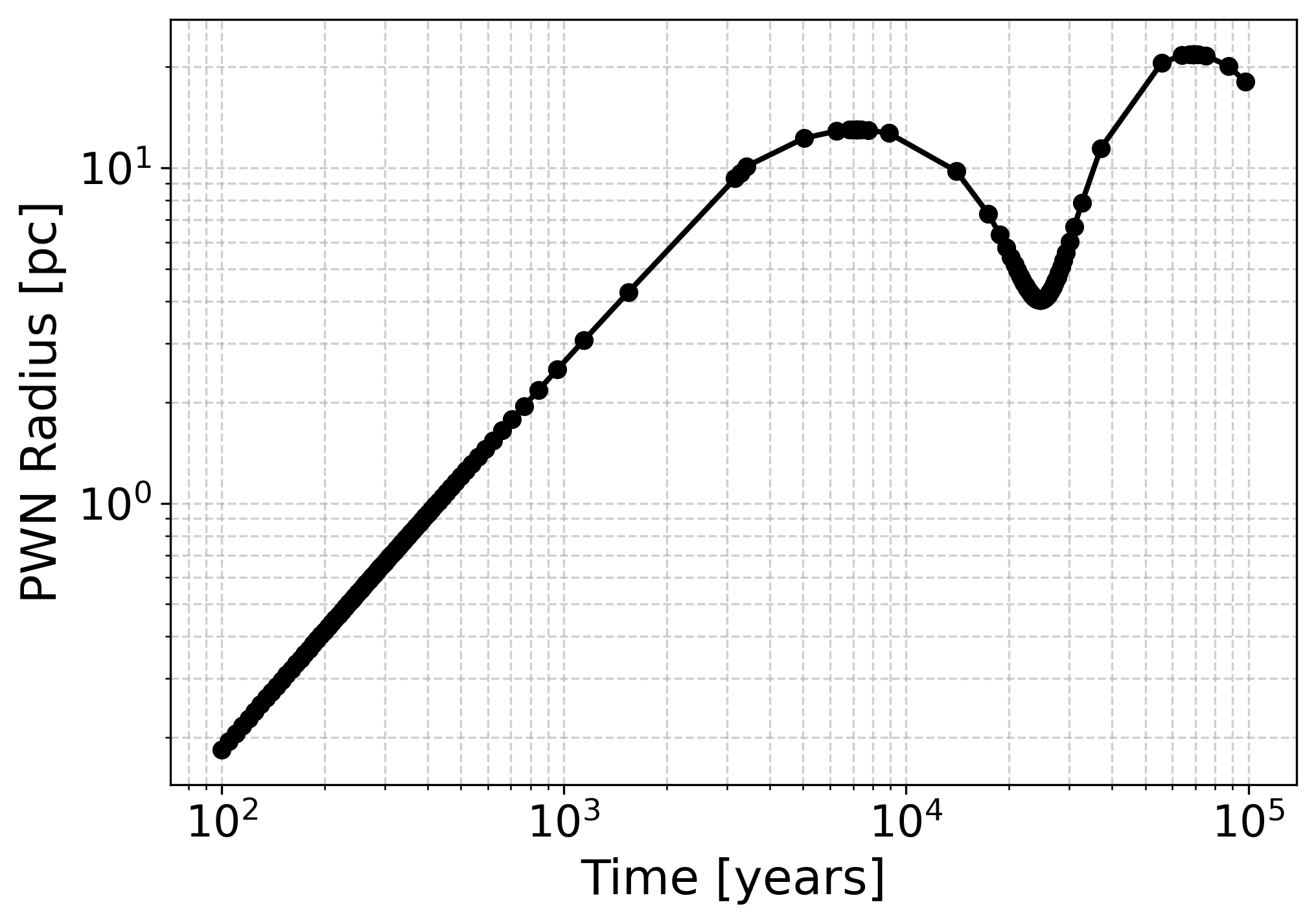}
    \includegraphics[width=0.475\textwidth]{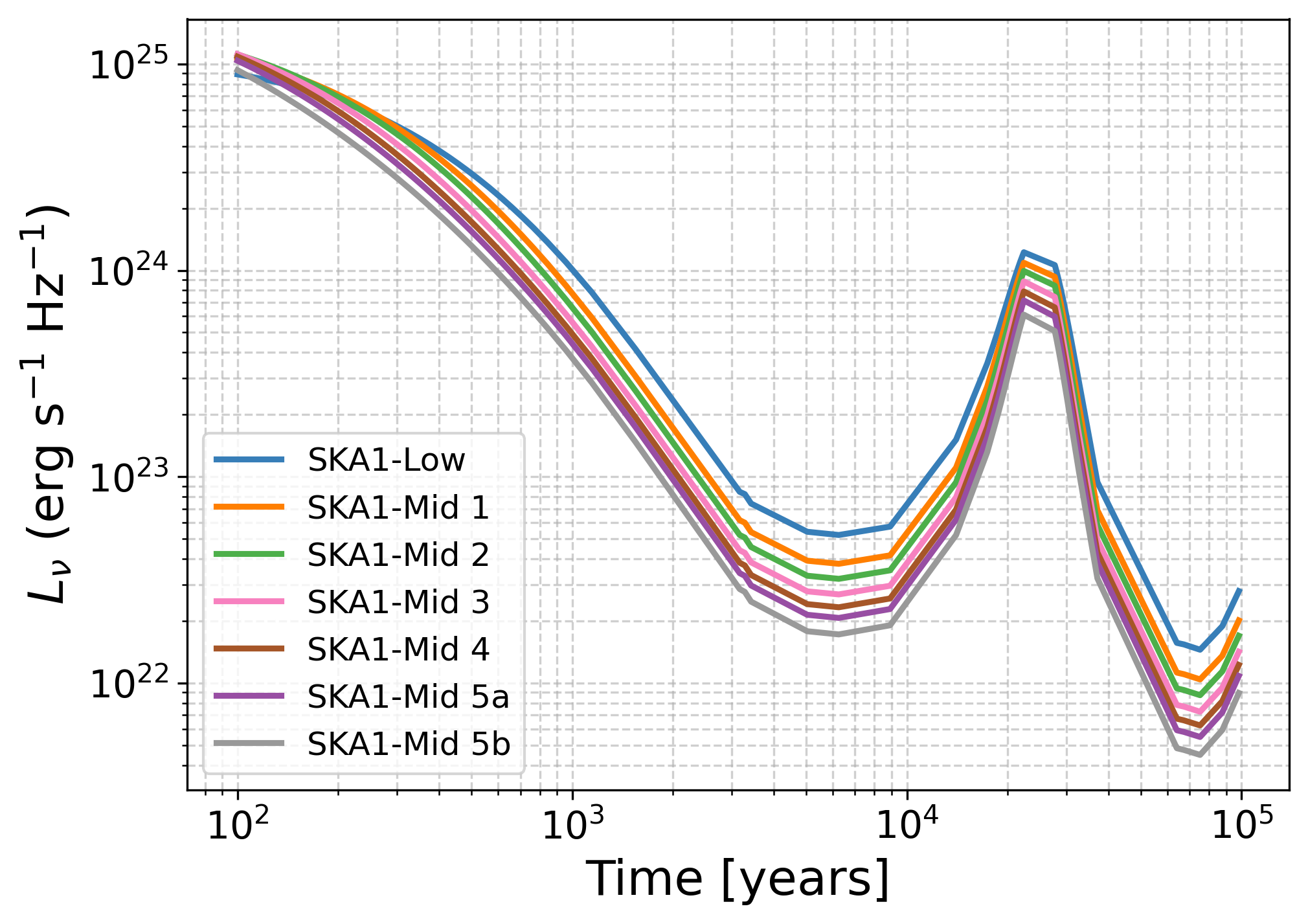}
    \includegraphics[width=0.475\textwidth]{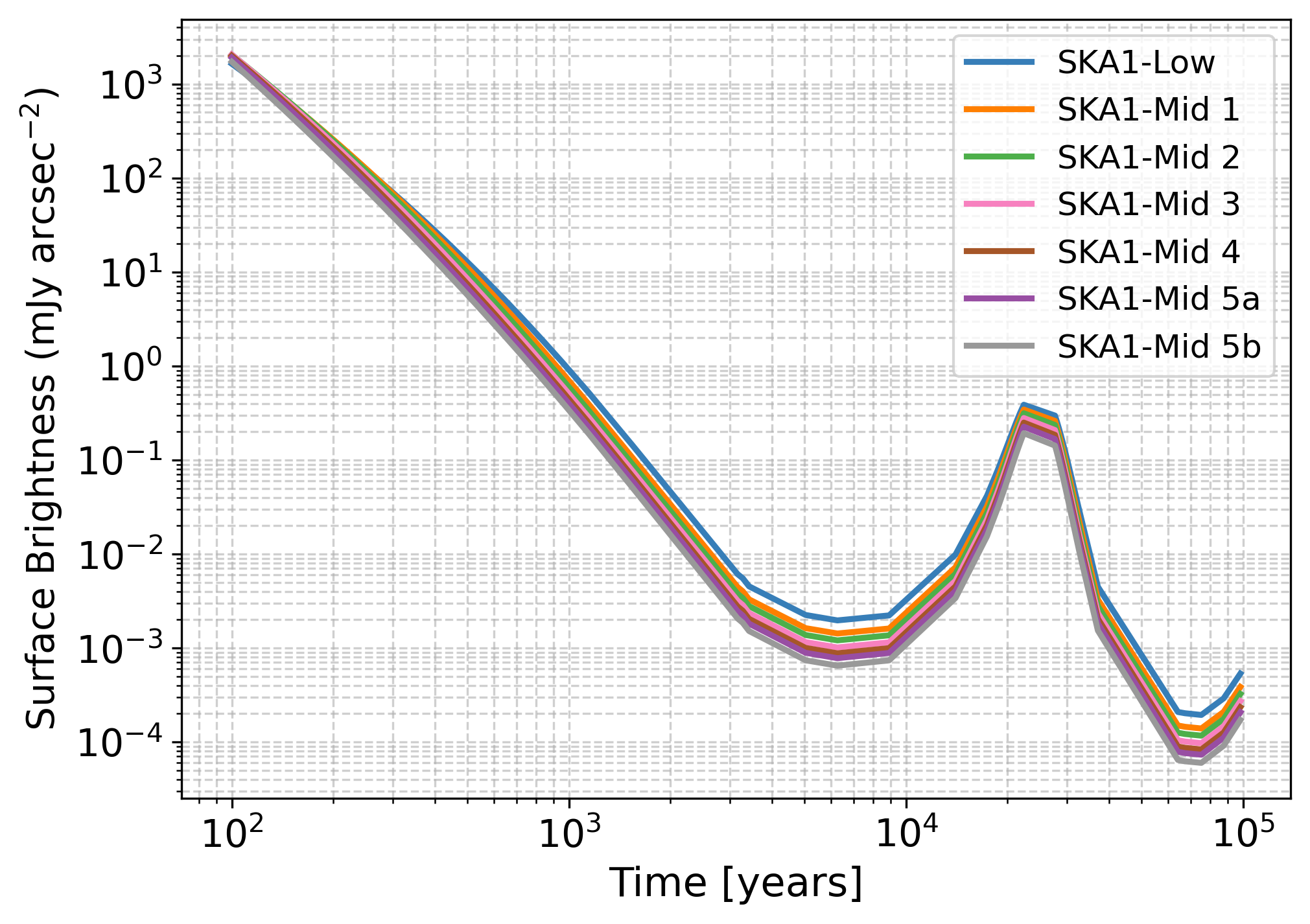}
    \includegraphics[width=0.475\textwidth]{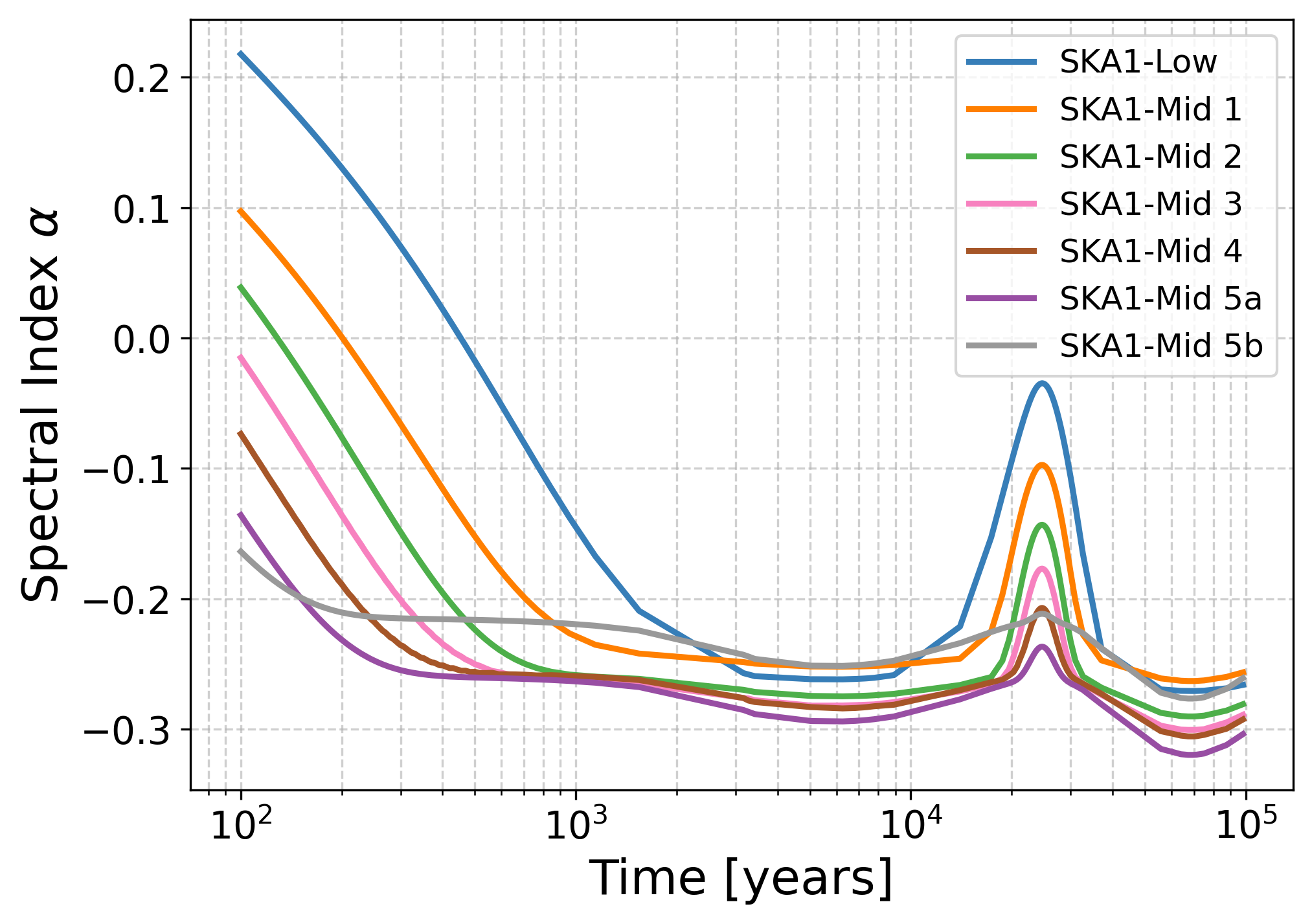}
    \caption{Evolution of the size ({\it upper left}), specific luminosity ({\it upper right}), surface brightness at a distance of 5 kpc ({\it lower left}), and spectral index of a PWN ({\it lower right}) inside a SNR predicted by the \citet{gelfand09} for the input parameters given in Table \ref{tab:model_parameters}.}
    \label{fig:radio_evol}
\end{figure}

Current theoretical models are able to predict the PWN's radio emission (e.g., \citealt{gelfand09}) when the pulsar is still within the SNR formed by its progenitor supernova (\S\ref{sec:evolution}).  While the evolution of a particular PWN is strongly dependent on the properties of the system (e.g., Table \ref{tab:model_parameters}), as shown in Figure \ref{fig:radio_evol} during this period the radius, specific luminosity, surface brightness, and radio spectral index $\alpha$ (flux density $S_\nu \propto \nu^\alpha$) changes significantly over time. However, throughout the evolution of the PWN it is expected that the spectral index $\alpha$ (flux density $S_\nu \propto \nu^\alpha$) of its radio emission changes between the SKA Low and SKA-Mid bands, with a decrease in $\alpha$ as frequency $\nu$ increases.  Furthermore, the compression of the PWN by the reverse shock inside the SNR results in an increase in luminosity,  surface brightness, and spectral index $\alpha$ -- with the change in spectral index most prominent, and occuring earlier, at lower frequencies. Therefore, significant spectral curvature observed in the range of frequencies observable by the SKA may indicate this interaction has begun.  Furthermore, the higher surface brightness inferred for the youngest and compressed PWN also suggests the planned surveys discussed above will preferentially detect sources in these two evolutionary phases.



The scientific investigations mentioned above requires measuring the spectrum and morphology of the radio emission from PWNe associated with a wide variety of NSs.  Since most NSs are located in the Galactic Plane, PWNe are often in fields with numerous, much brighter, objects where source confusion is a real and pressing concern.  Therefore, identifying PWNe requires images with the high dynamic range only achievable with the excellent {\it u-v} coverage promised by distribution of elements with the SKA.  The broadband coverage of the SKA will make it possible to produce such images at a wide range of frequencies, enabling the accurate measurements of the PWN's flux density across the radio band, as needed for the spectral studies above. 

As shown in Figure \ref{fig:radio_evol}, the radius of a PWN is expected to change by more than an order of magnitude over its lifetime.  Even at a distance of 10~kpc, only sub-arcminiute angular resolution is needed to resolve the bulk radio emission from a PWN.  However, the longest baselines of the SKA will provide the sub-arcsecond angular resolution needed to measure small-scale pulsar proper motion vectors, and to study small-scale structures within PWNe.  This capability will be especially important for comparing the morphology of PWNe in different wavebands.  Offsets between the peaks of X-ray and radio emission have been observed in several bow-shock pulsar wind nebulae, including J1509--5850~\citep{2008ApJ...684..542K}, G319.9--0.7~\citep{2010ApJ...712..596N}, the Lighthouse Nebula~\citep{2014A&A...562A.122P}, Potorro~\citep{lazarevic24}, and PSR J2030+4415~\citep{paredes25}. As discussed in previous sections, X-ray emission is generally believed to originate near the particle acceleration site. Therefore, the observed offsets between the X-ray and radio peaks suggest a gradual decrease in the radio surface brightness of the PWN tail toward the pulsar. This decline in radio brightness near the pulsar cannot be attributed to synchrotron cooling, as the cooling time for radio-emitting electrons is relatively long. As proposed by \citet{2008ApJ...684..542K}, explaining the absence of radio emission close to the pulsar may require assuming an increase in particle number density with distance from the pulsar.

Meanwhile, the short baselines allow detecting radio emission on the large angular scales.  This will be especially important for detecting radio emission from the large scales of the TeV halos discussed above -- whose number at the PeV energies can potentially be increased using SKA-Low \citep{Nelles01.2026.SKA}.  The interaction between the high-energy particles responsible for the observed $\gamma$-rays and the interstellar magnetic field will result in synchrotron radiation whose detection is vital for determining the spectrum of these $e^\pm$ as well as strength and structure of the local magnetic field.  Previous searches for this synchrotron emission at infrared (e.g., \citealt{hooper24}) and X-ray energies (e.g., \citep{khokhriakova24}) have been unsuccessful due to the complicated background and low surface brightness of this emission.  Due to the increased number of the radio-emitting particles, the surface brightness of the synchrotron emission from such sources is likely to be brightest at these wavebands.  With interferometric observations, it is possible to filter the diffuse Galactic background from these images.  Unfortunately, current interferometers are not sufficiently sensitive to emission originating from the $\gtrsim1-2^\circ$ angular size of the $\gamma$-ray emission detected from these sources -- an issue which will be rectified with the SKA.

The detection of polarized emission from PWNe is also of vital importance.  Not only do such observations directly probe the structure of the nebular magnetic field, recent studies by \citet{LP2016ApJ, LP2012ApJ} showed that the anisotropy and the fluctuation spectrum of turbulence can be determined by measuring the fluctuations in polarization of synchrotron radiation from a source.  Such studies would complement investigations of magnetic turbulence inside a PWN made using the spatial two-point correlation function of radio intensities \citep{ROY2009MNRAS, SJ2018MNRAS, VP2020MNRAS}.  For example, the SKA would be able to repeat, with improved sensitivity and angular resolution, the large scale radio polarization study of TeV halo presented by \citep{malik24} -- critical for determining if magnetic turbulence is i.ndeed responsible for the low diffusion coefficient inferred for such sources.  The capability of SKA telescope to measure the radio polarization of PWNe with high angular resolution and sensitivity is critical for understanding the nature of turbulence within these objects.

Lastly, a significant fraction of radio PWNe are currently unassociated with a pulsar.  While some of these could be attributed to beaming, the increased sensitivity of the SKA allows for the discovery of fainter pulsars associated with these objects \citep{Keane01.2026.SKA}.  Both, the current spin-down luminosity and characteristic age inferred from such detections are extremely helpful for studying the energetics of such objects.  Furthermore, the discovery of a radio pulsar within a PWN opens the possibility of conducting pulse-binning observations with long-baseline interferometers, allowing the pulsar’s position to be determined with sub-arcsecond precision. The polarization properties of the radio pulses can also be used to constrain the magnetic and rotational axes of these objects.  In parallel, deep, high-resolution radio observations can reveal the detailed structure of the radio tail and enable direct comparisons with X-ray morphology and pulsar geometry. Together, these observations are essential for advancing our understanding of collimated, magnetized outflows from pulsars.

While pointed observation will likely be needed to comprehensively study a large sample of PWNe, the surveys planned with the initial AA$^\star$ configuration of the SKA and -- eventually -- the complete SKA offer considerable opportunities for serendipitous discoveries.  As demonstrated by SKA pathfinders MeerKAT (e.g., \citealt{heywood22, motta23}) and ASKAP (e.g., \citealt{umana21}), surveys of the Galactic plane with the improved $u-v$ coverage and sensitivity offered by these new facilities are capable of significantly improving measurements of the radio properties of known PWNe and discovering new radio PWN candidates around known pulsars.  In addition, the success of these and other surveys in discovering supernova remnants (e.g., \citealt{anderson17, mantovanini24,  smeaton24}) suggests the SKA has the potential of discovering SNRs \citep{Ingallinera01.2026.SKA} around the considerable fraction of PWNe powered by young ($\lesssim10^4$~year) neutron stars currently without such associations.  Such detections would be very useful in determining the nature of the stellar progenitors of these objects, in addition to several of the investigations discussed in \S\ref{sec:open}.  These imaging surveys, coupled with the pulsar searches planned for these regions of the sky \citep{Keane01.2026.SKA}, have the potential of significantly increasing the number of radio PWNe associated with secure neutron star detections -- critical for answering the science questions discussed above.  The improvements in sensitivity, timing and polarization capabilities, and frequency coverage offered by the full SKA promise to further the advances made by AA$^\star$.  Furthermore, by virtue of the broadband emission of PWN, these SKA studies will have considerable synergies with new observing facilities at other wavebands on similar timescale, such as the Cerenkov Telescope Array (CTA) at $\gamma$-ray energies and forthcoming X-ray satellites.




\section{Conclusions}
PWNe are laboratories for studying the behavior of material at magnetic field strengths and particle energies rarely achieved in other astronomical sources.   Radio studies of PWNe are critical for understanding the production of particles in neutron star magnetospheres, the acceleration of $e^\pm$ to PeV energies, and the physics of ultra-relativistic outflows.  The improvements in image fidelity and sensitivity to pulsed, continuum, and polarized radio emission on both large and small angular scales offered by the SKA in both its initial phase and final configuration will allow detailed studies of the radio emission of a large and varied sample of these sources -- as needed to understand the fundamental physical processes operating in these high-energy sources.

\section*{Acknowledgments}
The authors would like to thank Bhal Chandra Joshi for useful comments.  Basic research at NYU Abu Dhabi is supported by the Crown Prince's Court of the Emirate of Abu Dhabi, as administered by Tamkeen.  JDG and MA are supported by NYUAD research grant AD022, and the Center for Astrophysics Space Science is support by a grant from the NYUAD Research Institute.  C.-Y. N. and Y. S. are supported by a GRF grant of the Hong Kong Government under HKU 17304524.

\bibliographystyle{abbrvnat-maxbibnames4}

\bibliography{chapter}

\end{document}